\documentclass[letterpaper]{article}
\usepackage{aaai2027}
\usepackage[hyphens]{url}
\usepackage{graphicx}
\usepackage{natbib}
\usepackage{caption}
\usepackage{amsmath}
\usepackage{amssymb}
\usepackage{subfig}
\usepackage{booktabs}
\usepackage{multirow}
\usepackage{makecell}
\usepackage{multicol}
\usepackage{xcolor}
\usepackage{tabularx}
\usepackage{array}

\title{HyWA: Architecture-Preserving Personalized Voice Activity Detection for Full-Duplex Voice Assistants}

\author{
    Hamed Jafarzadeh Asl\textsuperscript{\rm 1}\corresponding\equalcontrib,
    Amin Edraki\textsuperscript{\rm 1}\equalcontrib,
    Mahsa Ghazvini Nejad\textsuperscript{\rm 1}\equalcontrib,
    Masoud Asgharian\textsuperscript{\rm 2},
    Mohammadreza Sadeghi\textsuperscript{\rm 1},
    Yuanhao Yu\textsuperscript{\rm 1},
    Vahid Partovi Nia\textsuperscript{\rm 1,3}
}

\affiliations{
    \textsuperscript{\rm 1}Huawei Noah's Ark Lab, Canada~~~
    \textsuperscript{\rm 2}McGill University,  Canada~~~
    \textsuperscript{\rm 3}Polytechnique Montreal, Canada
    hamed.jafarzadeh.asl@h-partners.com
}

\begin{document}

\maketitle

\begin{abstract}
    Voice activity detection (VAD) serves as an early gate in voice-assistant pipelines for smart devices. Because conventional VADs respond to speech from any speaker, nearby conversations and residual assistant playback lead to unwanted triggers, degrade the user experience, and waste computational resources. Personalized voice activity detection (PVAD) addresses this limitation by detecting speech only from an enrolled target speaker. Existing PVAD methods typically incorporate speaker information into model inputs or hidden representations. These approaches require VAD architectural changes that increase engineering and requalification costs in deployment. We present HyWA, a hypernetwork-based weight-adaptation method that converts an established VAD into a PVAD, while preserving its acoustic interface and inference topology. HyWA generates speaker-conditioned weights once at enrollment and requires no per-user optimization. We apply HyWA to pretrained Alibaba's FSMN and NVIDIA's MarbleNet VAD models. Evaluations on synthetic and real-user test data show consistent improvements in target-speaker discrimination and false-positive suppression. In an integrated full-duplex barge-in pipeline, replacing generic VAD with HyWA-based PVAD reduces observed false-interruption detections from 88.9\% to only 9.9\%.
\end{abstract}

\vspace{-.7em}
\section{Introduction}
\label{sec:intro}

Voice activity detection (VAD) is the task of determining whether an audio signal contains human speech or is only non-speech background noise. It is a fundamental preprocessing step in many speech processing systems, including automatic speech recognition, speaker verification, and audio indexing \cite{ramirez2007voice}. Over the years, VAD methods have evolved from simple energy-based and statistical approaches towards more robust machine learning and deep learning techniques. Early methods mainly relied on hand-crafted features and thresholding rules, Gaussian mixture models, hidden Markov models, and other probabilistic frameworks. More recently, neural network-based VAD models have achieved significant improvements by learning discriminative features directly from data. Representative examples include MarbleNet~\cite{Jia_MarbleNet_ICASSP2021} and FSMN~\cite{Gao_FunASR_InterSpeech2023}, which exploit convolutional and feedforward sequential architectures.

\begin{figure}
    \centering
    \includegraphics[width=0.98\linewidth]{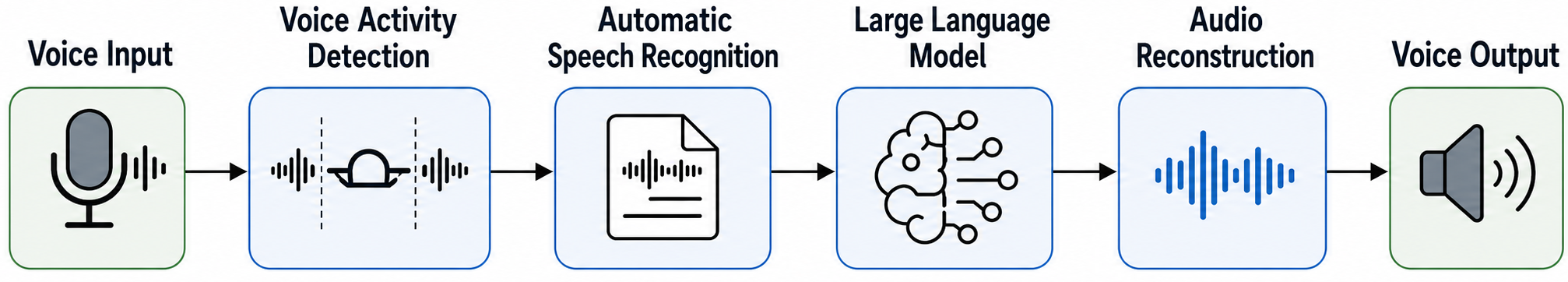}
    \caption{A typical voice assistant pipeline that includes voice activity detection (VAD) as the early gating module.}
    \label{fig:voice_assistant}
    \vspace{-1.3em}
\end{figure}

The new era of artificial intelligence is evolving with the rise of smart devices and ambient computing, where voice assistants have become one of the most natural interfaces between humans and technology. Leading platforms such as Google Assistant, Amazon Alexa, Microsoft Cortana, and Huawei Celia made voice interaction a mainstream part of consumer electronics. They help users set reminders, control smart homes, search for information, and interact with mobile services through simple spoken commands. In a typical voice assistant pipeline~(Figure~\ref{fig:voice_assistant}), VAD acts as the first gate that decides whether the incoming audio stream should be forwarded to more expensive downstream modules such as automatic speech recognition.
Conventional VAD systems operate at the 10--20~ms time frame level, and make binary \{``speech", ``non-speech"\} decisions for each incoming audio  frame~\cite{OURS_MLSP2025_tinynoiserobustvoiceactivity}.

Personalized VAD (PVAD) \cite{personal_vad_1, personal_vad_2} offers several practical advantages over conventional VAD. PVAD reduces false triggers caused by nearby conversations or overlapping speech, by focusing on the enrolled target speaker. It also reduces unnecessary computation by avoiding downstream processing when the input signal does not contain speech from the intended user.
PVAD models modify traditional VAD models to incorporate the speaker identity into the VAD processing pipeline~\cite{medennikov2020target}. They typically rely on a speaker embedding extracted from a short enrollment utterance of a target speaker. This embedding is then injected into the model through \emph{speaker conditioning}~\cite{wang2018voicefilter}.
Deploying such PVADs requires speaker-conditioning modules, architectural modifications, and retraining of the base VAD model~\cite{kang2023svvad, SSL-pvad-baseline}.
These modifications increase engineering effort and complicate their adoption in an established voice assistant pipeline.

In our deployment setting, the established acoustic interface, tensor shapes, and speech-time execution graph had to remain unchanged. However, enrollment-time processing and loading of user-specific parameter values are feasible. These limitations call for new methods that can turn a standard VAD into a personalized system with minimal changes to the original deployment flow.
We propose a new perspective for speaker conditioning that leverages a hypernetwork~\cite{ha2017hypernetworks} to personalize an existing VAD once at enrollment. Our hypernetwork is an auxiliary model that generates weight updates for an existing VAD model to adapt it for a target speaker. This approach preserves the core VAD architecture and minimizes the PVAD deployment effort in commercial scenarios by re-using the same VAD pipeline.

One important deployment scenario for PVAD is full-duplex voice interaction, in which a user may interrupt, or barge-in on a voice assistant while the assistant is producing speech. In this setting, the speech detector must preserve genuine target-user interruptions while rejecting residual assistant playback and other speech-like artifacts produced by the device and acoustic environment. This application therefore requires both reliable target-speaker detection and strong suppression of false interruptions. We first evaluate these general target-detection capabilities under controlled and real-user conditions, and subsequently examine the proposed PVAD within a fully integrated barge-in pipeline under deployment-specific failure cases.

\begin{figure}[!t]
    \centering
    \includegraphics[width=.95\columnwidth]{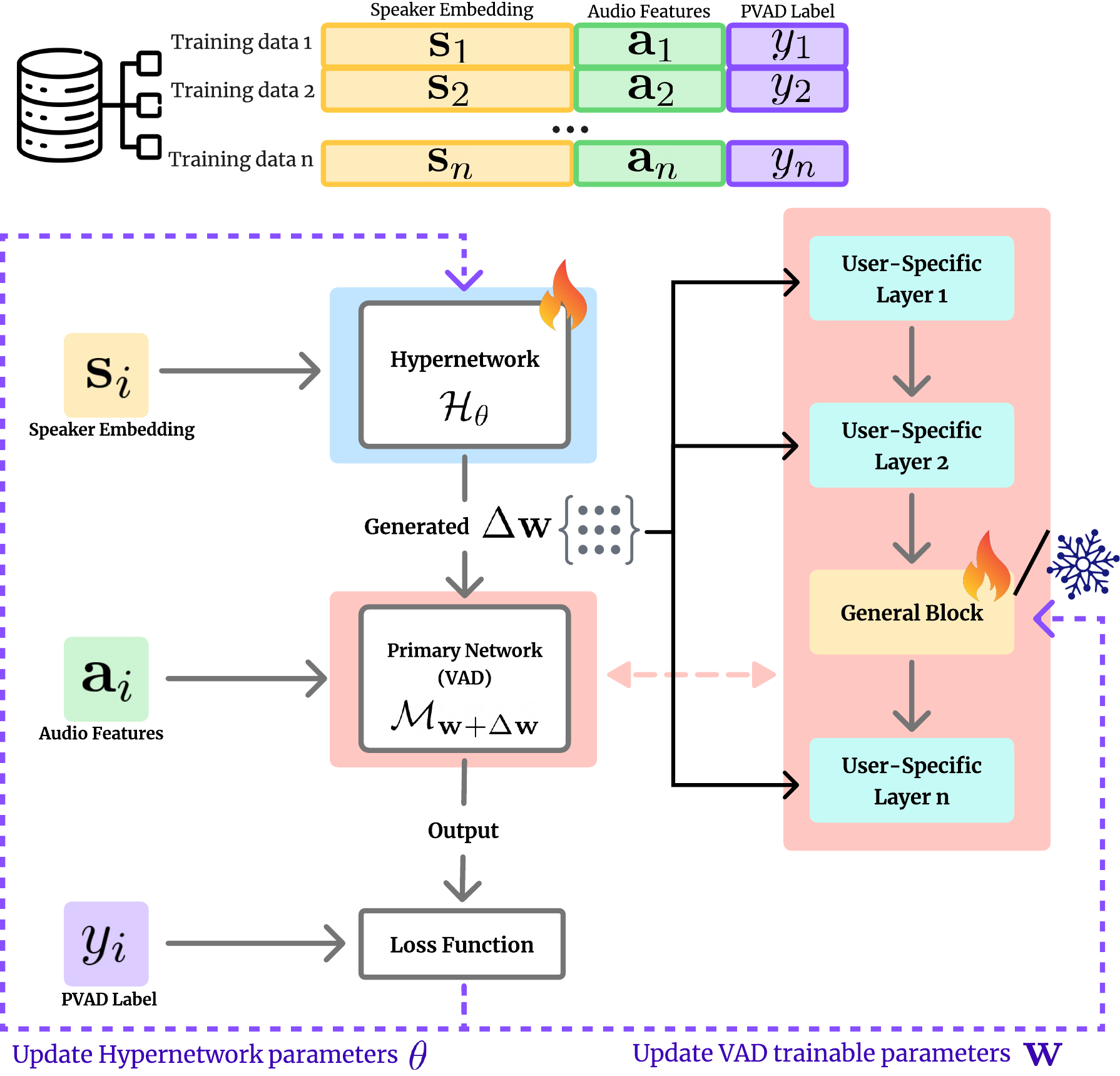}
    \caption{HyWA training pipeline for PVAD. The hypernetwork $\mathcal H_\theta$ generates user-specific residual updates $\Delta\mathbf w$ from speaker embedding $\mathbf s_i$. The resulting personalized VAD processes acoustic features $\mathbf a_i$ to predict PVAD label $y_i$.}
    \label{fig:training}
    \vspace{-1em}
\end{figure}

\section{Development}
\label{sec:method}

PVAD requires fusion of target-speaker enrollment information with acoustic features. HyWA moves this combination to enrollment time: a hypernetwork converts a speaker embedding into residual VAD weights, after which speech-time inference uses only acoustic features.

\subsection{Hypernetwork and Personalization Method}
\label{subsec:personaliztion}

PVADs commonly add user-specific features through concatenation, addition, multiplication, or feature-wise linear modulation (FiLM)~\cite{perez2018film,SSL-pvad-baseline-journalVersion,comparative-analysis-personalized-voice}. These approaches alter the input or add conditioning operations to the speech-time model.
Personalizing the weights of an existing VAD provides an architecture-preserving alternative. Instead of changing the input to obtain a PVAD, we generate the VAD weights for that specific user through a hypernetwork. A hypernetwork, introduced in~\cite{ha2017hypernetworks}, is a metamodel that generates the weights of a primary model using metadata. In HyWA, this metadata is a speaker embedding extracted from the target user's enrollment audio. Instead of direct personalization of primary VAD model parameters, the hypernetwork learns a mapping from this embedding to residual updates in the VAD weight space. This approach enables enrolling new users without per-user optimization.

Assume a VAD model is $\mathcal M_{\mathbf w}(\cdot)$ and the personalized VAD model, PVAD, for user $k$ is $\mathcal M_{\mathbf w_k}(\cdot)$.
We implicitly assume a user $k$ has a speech characteristic vector $\mathbf s_k$, also referred to as speaker embedding.
Our hypernetwork aims to generate user-specific weights conditioned on $\mathbf s_k$.
We reparameterize the personalized weights as $\mathbf w_k = \mathbf w + \Delta \mathbf w_k$, where $\Delta \mathbf w_k$ captures a user-specific residual update over the shared VAD parameter $\mathbf w$.
In HyWA, $\Delta \mathbf w_k$ is generated in a single forward pass of a hypernetwork conditioned on $\mathbf s_k$. This mechanism is particularly advantageous for personalization, because hypernetwork generates user-specific weights by only modifying certain layers of the primary VAD model~\cite{chauhan2023hypernetworks}. Training is performed over a set of users $i \in \{1,\ldots,n\}$ with speaker embeddings~$\mathbf s_i$.

Unlike feature-injection approaches, HyWA adds no speaker input or conditioning module to the speech-time graph. Appendix~\ref{app:hywa_proof_of_concept} compares weight conditioning with common alternatives in a controlled model. The primary application rationale, however, is satisfying the fixed-topology deployment constraint rather than claiming universal superiority over all PVAD architectures.

\subsection{Training}
\label{subsec:training_pipeline}

Training of a VAD model $ y_t \!\sim\! \mathcal M_{\mathbf w}(\mathbf a_t)$ optimizes a loss over acoustic input $\mathbf a_t$ and its binary speech label $y_t\!\in\!\{\text{``speech''}, \text{``non-speech''}\}$ at each timestamp $t$. In the sequel, we drop timestamp $t$ and user $k$ indices when there is no danger of notation confusion. 

The training pipeline is visualized in Figure~\ref{fig:training}. The backbone parameters $\mathbf w$ may be learned jointly with the hypernetwork when training from scratch, or held fixed, i.e. frozen, when adapting a pretrained VAD. Given an enrollment embedding $\mathbf s$, the hypernetwork generates residuals $\Delta\mathbf w=\mathcal H_\theta(\mathbf s)$, which are added to selected backbone parameters. Then, PVAD outputs the labels $y$ based on the acoustic features $\mathbf a$.
Each training example consists of a speaker embedding $\mathbf s$, acoustic features $\mathbf a$, and a PVAD label $y$. For target-speaker detection, $y$ may be ternary, from \{``non-speech~(ns)'', ``target-speaker speech~(tss)'', ``non-target-speaker speech~(ntss)''\}. To preserve the conventional binary VAD interface in our pretrained-backbone experiments, we assign ``tss'' to $y=1$ and \{``ntss'', ``ns''\} to $y=0$.

A cross-entropy objective is used to optimize the hypernetwork parameters $\theta$. In the from-scratch proof-of-concept experiment, the backbone parameters $\mathbf w$ are additionally optimized jointly with $\theta$. Because $\Delta\mathbf w$ is generated by $\mathcal H_\theta$, it is not an independently optimized parameter.

\begin{figure}[t]
  \centering
  \subfloat[Enrollment and Deployment]{%
    \fbox{\includegraphics[width=.49\linewidth]{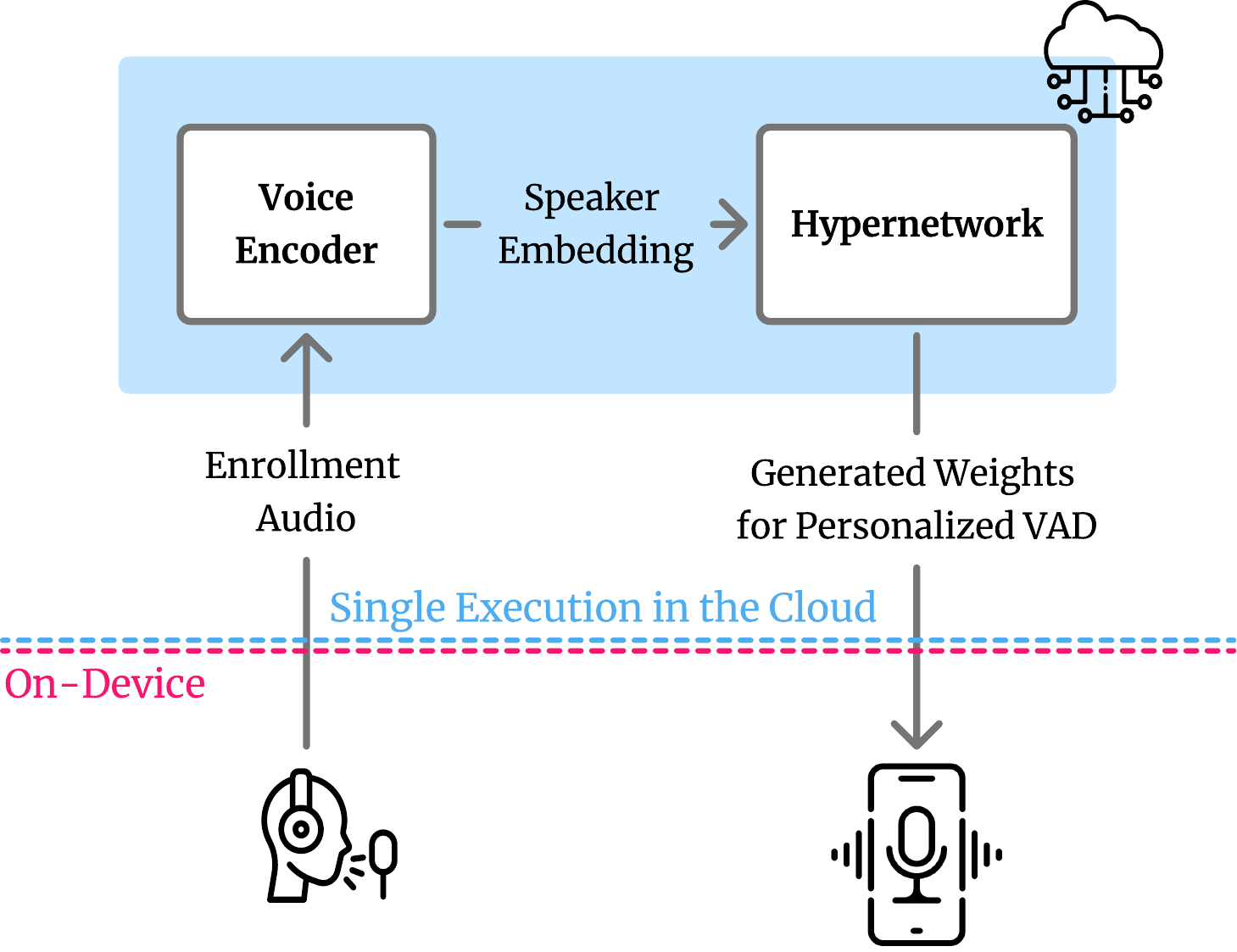}}%
    \label{fig:enrollment}
  }\hfill
  \subfloat[Usage]{%
    \fbox{\includegraphics[width=.42\linewidth]{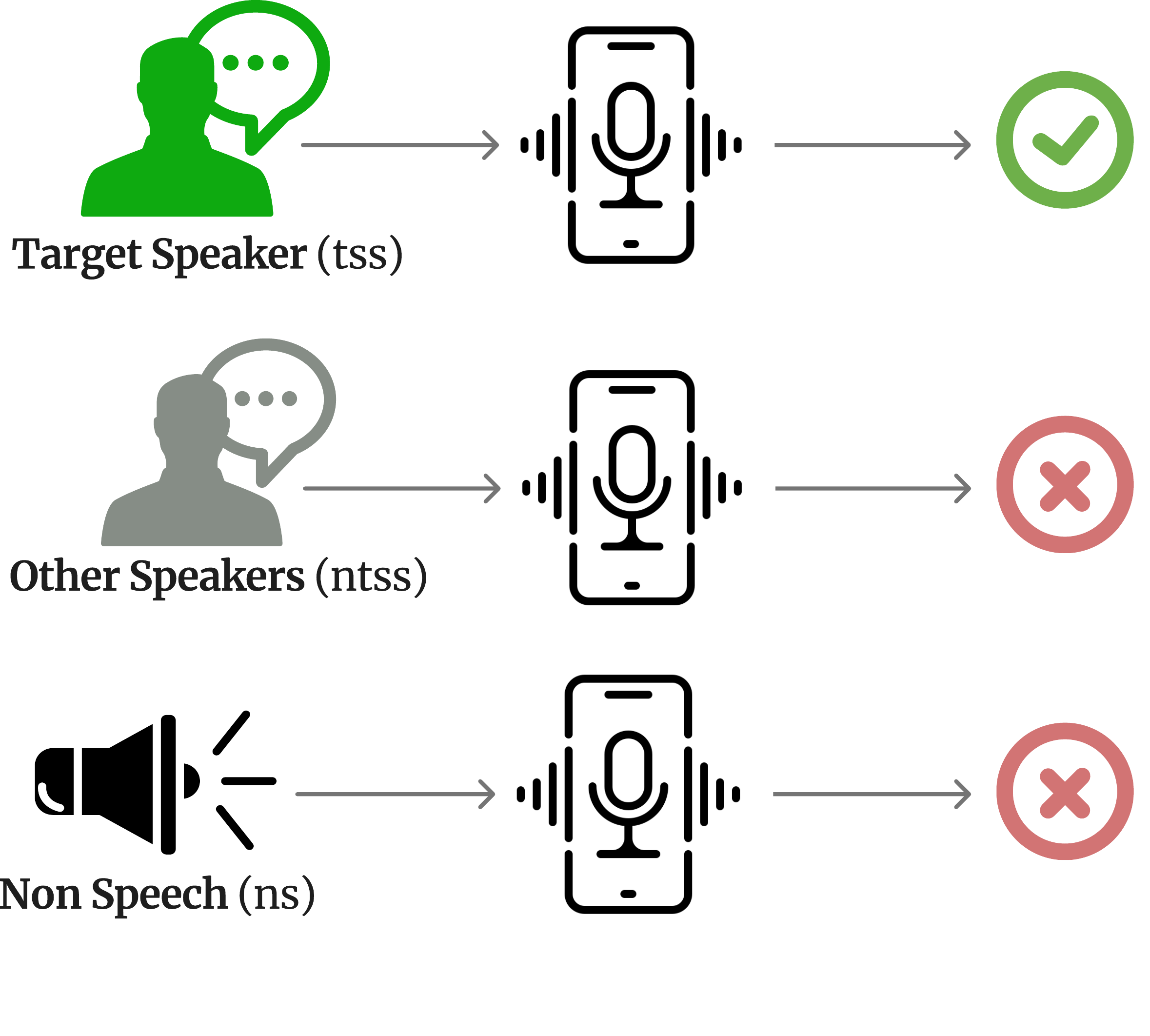}}%
    \label{fig:usage}
  }
  \caption{HyWA-PVAD inference pipeline. (a) Enrollment and deployment are executed once per user through cloud-device communication. (b) Speech-time usage is executed on the device after the personalized weights have been loaded.}

  \label{fig:inference}
  \vspace{-0.6em}
\end{figure}

\subsection{Inference}
\label{subsec:inference_pipeline}

After training the hypernetwork, the steps to use the PVAD model for an enrolled user include
\begin{itemize}
    \item[i)] \emph{Enrollment:} feeding speaker embedding $\mathbf s$ into the hypernetwork $\mathcal H_\theta$ to generate  personalized $\Delta \mathbf w$,
    \item[ii)] \emph{Deployment:} feeding $\Delta \mathbf w$  into $\mathcal M$ to obtain a PVAD model $\mathcal M_{\mathbf w + \Delta  \mathbf w}$, see Figure~\ref{fig:enrollment},
    \item[iii)] \emph{Usage:} feeding acoustic features $\mathbf a$ at every timestamp into $\mathcal M_{\mathbf w + \Delta \mathbf w}$ to obtain personalized labels~$y$, see Figure~\ref{fig:usage}. 
\end{itemize}

Figure~\ref{fig:enrollment} illustrates one cloud-device deployment configuration. For privacy-sensitive applications, the voice encoder can run on-device so that only the resulting speaker embedding is transmitted to the hypernetwork. Once the residuals are added, $\mathcal M_{\mathbf w+\Delta\mathbf w}$ has the same layers, tensor shapes, acoustic input, number of active backbone parameters, and speech-time operations as $\mathcal M_{\mathbf w}$. The hypernetwork is executed only during enrollment and is absent from frame-level inference. Thus, HyWA does not increase the backbone's speech-time operation count. The personalized model therefore requires only acoustic features $\mathbf a$ during speech-time inference; see Figure~\ref{fig:usage}.
We consider a personal-device setting with one enrolled user, as in smartphone voice assistants. The active speaker profile is therefore known before speech-time inference.

\section{Pre-Deployment}
\label{sec:predeployment}

We evaluate HyWA in three stages.
First, we provide a proof of concept for our approach. We apply HyWA to a compact LSTM architecture with approximately 85K parameters. This architecture is trained jointly with the hypernetwork to compare HyWA with common speaker-conditioning methods. This PVAD is evaluated in two task formulations: a binary variant used as the Base reference in Table~\ref{tab:librispeech_results}, and a ternary variant used for the conditioning comparison in Appendix~\ref{app:hywa_proof_of_concept}.
This experiment establishes the feasibility of generating speaker-conditioned VAD weights under a controlled setting. The hypernetwork architecture, compact LSTM model, conditioning baselines, and results are reported in Appendix~\ref{app:hywa_proof_of_concept}.

Second, we personalize pretrained VAD backbones, Alibaba's FSMN and NVIDIA's MarbleNet, in a pre-deployment setting. This experiment represents a feasible integration path of a HyWA-PVAD into a voice assistant pipeline.
Finally, we evaluate our PVAD in an integrated full-duplex pipeline as the ultimate scenario before product handoff.

\subsection{Model Setting}
\label{sec:predeployment_models}

Pre-deployment experiments apply HyWA to pretrained FSMN and MarbleNet VAD backbones to obtain FSMN-PVAD and MarbleNet-PVAD. For each backbone, the pretrained parameters remain frozen, while a backbone-specific hypernetwork maps an enrollment speaker embedding to residual updates $\Delta\mathbf{w}_k$ for a selected subset of VAD parameters.
The personalized parameters are formed as $\mathbf{w}_k\!=\!\mathbf{w}\!+\!\Delta\mathbf{w}_k$ and loaded into the original backbone. The resulting PVAD retains the backbone's inference pipeline. After training the hypernetwork, enrolling a new user requires generating and loading a new set of residual updates rather than optimizing the VAD for that user.

FSMN-PVAD and MarbleNet-PVAD both use 256-dimensional, $\ell_2$-normalized enrollment embeddings, generated by Resemblyzer VoiceEncoder~\cite{icassp2018_resemblyzerpaper}. The two systems employ distinct hypernetworks, while keeping the pretrained VAD parameters frozen. Detailed configurations are provided in Appendix~\ref{app:predeployment_configs}. Overall, $\approx$165K and $\approx$80K user-specific parameters are generated for FSMN-PVAD and MarbleNet-PVAD, respectively.

\subsection{Datasets}
\label{sec:data}

The LibriSpeech and real-user datasets serve complementary but distinct roles in our experiments. We use LibriSpeech to train and develop the HyWA personalization components. The pretrained FSMN and MarbleNet VAD backbones remain frozen throughout training.
Before deployment, we evaluate the proposed approach in a controlled setting using the test subset of LibriSpeech. This benchmark evaluation measures target-speaker discrimination under standardized acoustic conditions.

The real-user data is reserved exclusively for pre-deployment evaluation under conditions representative of the intended voice-assistant application. No real-user recording is used for training or optimization of the VAD backbones, hypernetworks, or speaker encoder. The corpus introduces application-relevant speaker variability, background noise, and device-generated speech echo that are not fully represented by the LibriSpeech benchmark.
Together, these evaluations examine whether HyWA, trained on LibriSpeech, generalizes to real-world voice-assistant conditions in terms of target-speaker discrimination and operating performance.

\subsubsection{LibriSpeech Benchmark.}
\label{sec:librispeechdata}

We construct a simulated multi-speaker benchmark from the LibriSpeech ``train-clean-100'', ``dev-clean'', and ``test-clean'' subsets, following the mixture-generation procedure of~\cite{SSL-pvad-baseline}. The three subsets are used for training, development, and final testing, respectively, with disjoint speakers across the splits. Each example is formed by concatenating one to three utterances, with one of the participating speakers designated as the enrollment target. Forced alignment of the transcripts provides frame-level speech boundaries, and the LibriSpeech metadata provides the corresponding speaker identities~\cite{panayotov_Librispeech_2015,mcauliffe2017montreal}.

We apply multistyle training~\cite{prabhavalkar2015automatic} by augmenting the speech with noise from MUSAN~\cite{musan2015}.
Training noise is drawn exclusively from the free-sound subset at SNRs from $-5$ to $20$~dB in $5$~dB increments. Evaluation is conducted under three conditions: clean speech, seen noise drawn from the same free-sound collection, and unseen noise drawn from the sound-bible collection, which is excluded from training. The noisy evaluation sets use the same SNR range and sampling procedure as the training mixtures. Recorded room impulse responses are additionally applied during augmentation~\cite{ko2017study}.

The main benchmark evaluation uses binary target-presence labels consistent with the intended VAD interface: ``target-speaker speech~(tss)'' is positive, whereas ``non-target-speaker speech~(ntss)'' and ``non-speech~(ns)'' are negative.
The ternary HyWA proof of concept is described in Appendix~\ref{app:hywa_proof_of_concept}.
The data-preparation code will be released as part of the PVAD framework.

\subsubsection{Real-User Data.}
\label{sec:userdata}

The application-domain test corpus contains 1,560 two-channel recordings, each 10~seconds in duration, collected under real-world voice-assistant usage scenarios. It includes 27 users, not seen during the training, and 26 source/noise groups, each containing 60 recordings: 20 under high-, mid-, and low-SNR conditions. For enrollment, each of the 27 users provides 10 recordings. The raw enrollment duration averages 11.37~s per user, with a range of 8.59--16.28~s.
Recordings were captured using a custom mobile handset application on Huawei HarmonyOS. Participants were shown a short phrase and a visual cue indicating when to speak. The cue timing was recorded and used to initialize the VAD labels, which were independently corrected by two annotators. Samples with onset or offset disagreements greater than 100~ms were excluded; otherwise, the final boundaries were obtained by averaging the two annotations.

The corpus covers realistic ambient conditions, including in-car, television, office, music, street, café, and public-transit noise. It also includes device-generated speech echo, such as voice-assistant and phone-call playback, which is particularly challenging for VAD because it can resemble human speech. Echo signals were recorded separately in an acoustic studio and later added to the recording mixtures.
The target SNR ranges were greater than 15 dB, 5 to 10 dB, and -5 to 0 dB for the high-, mid-, and low-SNR conditions, respectively. Since the effective SNR depends on user speaking level and device distance, noise levels were selected to achieve these ranges on average based on representative clean recordings.

\subsection{Evaluation Metrics}
\label{sec:predeployment_metrics}

The FSMN-PVAD and MarbleNet-PVAD are evaluated on binary target-presence tasks. We report the area under the receiver operating characteristic curve (AUROC) and target-class average precision (AP) as threshold-independent measures of target-speaker discrimination.
For operating-point evaluation, models are calibrated independently using the \emph{development groups}. We use a target development-set frame-level false-positive rate (FPR) of 0.2 as a fixed analysis operating point. This threshold is selected before evaluation on the \emph{test groups} to compare models at a common false-positive tolerance. This value is an evaluation protocol rather than a final product threshold, which must be selected jointly with the upstream and downstream components during target-device validation.
LibriSpeech separates development and test groups with the ``dev-clean'' and ``test-clean'' subsets. For the real-user corpus, we divide the 26 source/noise groups into 8 development and 18 test groups. Development groups are not used to train or update the VAD, PVAD, speaker encoder, or hypernetwork. Each selected threshold is fixed before test evaluation; because calibration uses development data, the achieved test-set FPR is not constrained to 0.2.
Appendix~\ref{app:fpr_sensitivity} repeats this protocol at target development FPRs of 0.05, 0.10, and 0.20 without retraining any model.

We report balanced accuracy (BA) at the calibrated operating point, defined as
\[
\mathrm{BA}=\frac{\mathrm{TPR}+\mathrm{TNR}}{2},
\]
where TPR and TNR denote true-positive rate and true-negative rate, respectively. We also report false-positive seconds per target-absent hour (FP s/h), defined as
\[
\mathrm{FP~s/h}
=
3600\,
\frac{D_{\mathrm{FP}}}
     {D_{\mathrm{target\text{-}absent}}},
\]
where \(D_{\mathrm{FP}}\) is the total predicted-positive duration during target-absent periods and \(D_{\mathrm{target\text{-}absent}}\) is the total target-absent duration. Lower FP s/h indicates less erroneous activation during non-target speech or non-speech. Together with TPR, this metric characterizes the operating trade-off between retaining enrolled-user speech and suppressing activation when the target speaker is absent.

Event-level performance is evaluated on each model's native score timeline. Predicted-positive regions are formed after merging gaps shorter than 100~ms. A target event is detected when at least 100~ms of predicted-positive time overlaps its annotated interval, and target-event recall is the proportion of annotated target events detected. A predicted-positive region that overlaps no target interval is counted as a false activation; false activations per hour are normalized by total recording time.

For paired PVAD--VAD comparisons in the real-user evaluation, we define $\Delta=\mathrm{PVAD}-\mathrm{VAD}$ and report point-wise 95\% percentile-bootstrap intervals over the held-out test groups. Appendix~\ref{app:event_bootstrap} provides the full procedure.

\begin{table*}[t]
\centering
\small
\setlength{\tabcolsep}{3pt}
\caption{Target-speaker evaluation on LibriSpeech. Values are mean (standard deviation) over \(N\!=\!5\) training seeds. Base is a binary version of the compact architecture trained for target presence. MarbleNet and FSMN denote the pretrained PVADs.}
\label{tab:librispeech_results}
\begin{tabular}{lccc ccc ccc}
\toprule
& \multicolumn{3}{c}{AUROC (\%) $\uparrow$}
& \multicolumn{3}{c}{AP (\%) $\uparrow$}
& \multicolumn{3}{c}{BA (\%) $\uparrow$} \\
\cmidrule(lr){2-4}
\cmidrule(lr){5-7}
\cmidrule(lr){8-10}
{Noise Condition}
& Base & MarbleNet & FSMN
& Base & MarbleNet & FSMN
& Base & MarbleNet & FSMN \\
\midrule
Clean
& 93.5 (0.6) & 91.9 (0.3) & {94.3} (0.2)
& 89.9 (0.7) & 86.1 (0.8) & {90.8} (0.3)
& 86.7 (0.7) & 85.0 (0.2) & {87.7} (0.4) \\
Seen (free-sound)
& 91.3 (0.4) & 89.5 (0.2) & {91.4} (0.4)
& 86.6 (0.5) & 82.5 (0.9) & {86.8} (0.5)
& 83.9 (0.4) & 82.5 (0.3) & {84.2} (0.5) \\
Unseen (sound-bible)
& 90.8 (0.8) & 89.6 (0.3) & {90.9} (0.5)
& 86.0 (1.0) & 82.9 (0.9) & {86.0} (0.6)
& 83.3 (0.9) & 82.4 (0.3) & {83.6} (0.6) \\
\bottomrule
\end{tabular}
\end{table*}

\section{Evaluation}
\label{sec:eval}

The intended application is target-selective gating in voice assistants, where a pretrained speaker-agnostic VAD is already part of the speech-processing pipeline. We investigate whether HyWA can convert pretrained VAD backbones into PVADs through enrollment-conditioned weight updates, while preserving each backbone's inference topology.

\subsection{LibriSpeech Benchmark}
\label{sec:librispeech_evaluation}

We first evaluate the two pretrained PVADs on the binary LibriSpeech benchmark under clean, seen-noise, and unseen-noise conditions. The binary compact model (base) is included in Table~\ref{tab:librispeech_results} only as a proof-of-concept reference; its ternary conditioning comparison appears in Appendix~\ref{app:hywa_proof_of_concept}. Because the three systems use different backbones and training procedures, this experiment evaluates target-speaker discrimination under common test conditions but does not isolate backbone or training effects.
Table~\ref{tab:librispeech_results} shows that all three systems maintain target-speaker discrimination across the evaluated acoustic conditions. Performance decreases under seen and unseen noise relative to clean speech. FSMN-PVAD obtains the highest AUROC, AP, and BA in all conditions.

\subsection{Real-User Data}
\label{sec:real_user_evaluation}

The real-user corpus provides the primary pre-deployment comparison. Each PVAD is evaluated against its matched speaker-agnostic VAD under the target-presence objective. Frame-level scores and annotations are mapped to a common 20-ms grid to prevent differences in native output rate from affecting the comparison.

\subsubsection{Target-Speaker Discrimination.}
To evaluate whether HyWA improves target-speaker discrimination in application-domain conditions, we compare each personalized model with its matched speaker-agnostic VAD on the common 20-ms test grid. Table~\ref{tab:application_common_grid} reports corpus-level AUROC, target AP, and BA computed after concatenating all retained test frames. PVAD standard deviations quantify training-seed variability.
FSMN-PVAD improves AUROC, target AP, and BA over FSMN-VAD by 10.3, 18.7, and 7.1 percentage points, respectively. MarbleNet-PVAD improves the corresponding metrics by 3.3, 7.8, and 2.9 percentage points relative to MarbleNet-VAD.

\begin{table}[t]
  \centering
  \caption{Target-presence evaluation on real-user data. VAD is reported as a fixed reference, while PVAD values are five-seed means with standard deviations.}
  \label{tab:application_common_grid}
  \small
  \setlength{\tabcolsep}{5pt}
  \begin{tabular}{llrr}
    \toprule
    Backbone & Measure & VAD & PVAD \\
    \midrule
    FSMN
      & AUROC (\%) $\uparrow$
      & 73.1
      & {83.4} (0.9) \\
      & AP (\%) $\uparrow$
      & 48.1
      & {66.8} (1.4) \\
      & BA (\%) $\uparrow$
      & 67.9
      & {75.0} (1.0) \\
    \midrule
    MarbleNet
      & AUROC (\%) $\uparrow$
      & 74.8
      & {78.1} (0.3) \\
      & AP (\%) $\uparrow$
      & 49.8
      & {57.6} (1.7) \\
      & BA (\%) $\uparrow$
      & 68.4
      & {71.3} (0.3) \\
    \bottomrule
  \end{tabular}
\end{table}

\subsubsection{Frame-level Performance.}

Table~\ref{tab:application_fp_reduction} reports frame-level performance at the development-calibrated $\text{FPR} = 0.2$ analysis operating point. Personalization reduces FP duration by $111.9$~s per target-absent hour for FSMN and $114.5$~s for MarbleNet.
Using $\Delta=\mathrm{PVAD}-\mathrm{VAD}$, the paired 95\% confidence intervals (CIs) are $[-197.7,-19.7]$ for FSMN and $[-192.9,-34.8]$ for MarbleNet. Both intervals exclude zero, indicating statistically significant reductions.
Target-speech TPR increases by 21.2 percentage points for FSMN (95\% interval $[17.0,25.5]$) and 4.3 percentage points for MarbleNet (95\% interval $[0.1,8.5]$); both intervals exclude zero.

\begin{table}[t]
  \centering
  \small
  \setlength{\tabcolsep}{4pt}
  \caption{Frame-level performance on the real-user test groups at the selected thresholds. Results are pooled across groups; VAD is a fixed reference, whereas PVAD point estimates are averaged over five seeds. Parentheses report standard errors.}
  \label{tab:application_fp_reduction}
  \begin{tabular}{@{}lcccc@{}}
    \toprule
    Backbone
      & \multicolumn{2}{c}{FP s/h $\downarrow$}
      & \multicolumn{2}{c}{TPR (\%) $\uparrow$} \\
    \cmidrule(lr){2-3}\cmidrule(lr){4-5}
      & VAD & PVAD
      & VAD & PVAD \\
    \midrule
    FSMN
      & 808.6 (23.4) & {696.6} (38.9)
      & 46.6 (1.1) & {67.8} (1.8) \\
    MarbleNet
      & 771.5 (28.3) & {657.0} (20.6)
      & 49.1 (1.6) & 53.4 (1.2) \\
    \bottomrule
  \end{tabular}
  \vspace{-0.7em}
\end{table}

\subsubsection{Event-Level Performance}

False activations per hour and FP duration measure different error properties. FP duration accumulates erroneous target-positive time and normalizes it by target-absent duration. In contrast, a false activation is a distinct predicted-positive region that overlaps no target event after gaps shorter than 100~ms are merged. Its frequency is normalized by total recording time. Event metrics are computed on each model's native valid score timeline using the same frozen $\text{FPR} = 0.2$ thresholds as the frame-level analysis. Consequently, the values in Tables~\ref{tab:application_fp_reduction} and \ref{tab:application_event_results} are neither numerically comparable nor expected to be equal.

\begin{table}[t]
  \centering
  \small
  \setlength{\tabcolsep}{4pt}
  \caption{Event-level performance on the real-user test groups at the selected thresholds. Results are pooled across groups; VAD is a fixed reference, whereas PVAD point estimates are averaged over five seeds. Parentheses report standard errors.}
  \label{tab:application_event_results}
  \begin{tabular}{@{}lcccc@{}}
    \toprule
    Backbone
      & \multicolumn{2}{c}{
          \shortstack{False activations per\\recording hour $\downarrow$}
        }
      & \multicolumn{2}{c}{
          \shortstack{Target-event\\recall (\%) $\uparrow$}
        } \\
    \cmidrule(lr){2-3}
    \cmidrule(lr){4-5}
      & VAD
      & PVAD
      & VAD
      & PVAD \\
    \midrule
    FSMN
      & 1759.0 (56.9)
      & 1170.4 (25.5)
      & 89.7 (0.9)
      & 94.8 (1.0) \\
    MarbleNet
      & 382.0 (14.3)
      & 418.3 (12.5)
      & 73.3 (1.9)
      & 74.1 (0.8) \\
    \bottomrule
  \end{tabular}
\end{table}

Using $\Delta=\mathrm{PVAD}\!-\!\mathrm{VAD}$, FSMN false-activation difference is $-588.6$ regions per recording hour, with a 95\% paired CI of $[-694.1,-481.5]$. It also improves target-event recall by $5.1$ percentage points, with an interval of $[3.5,6.9]$. Both effects are statistically significant.
The corresponding MarbleNet differences are $+36.3$ false activations per recording hour and $0.8$ percentage points in target-event recall. The corresponding intervals, $[-1.5,75.3]$ and $[-3.5,5.3]$, both include zero.

The event-level and frame-level findings are complementary. Both PVADs reduce the cumulative duration of false-positive predictions, but only FSMN-PVAD also produces a statistically significant reduction in the number of distinct false activations. A reduction in erroneous activation time therefore does not necessarily imply fewer activation events.
The two personalized backbones exhibit different operating trade-offs. MarbleNet-PVAD produces lower absolute FP duration and fewer false-activation events, whereas FSMN-PVAD achieves stronger target-speaker discrimination and substantially higher frame- and event-level recall.
Given the importance of preserving genuine user interruptions in the intended barge-in application, we advance FSMN-PVAD as the primary candidate along the path to deployment.

\section{Path to Deployment and Product Handoff}
\label{sec:path}

\begin{table}[t]
    \centering
    \small
    \setlength{\tabcolsep}{4pt}
        \caption{False interruption detections on the target-absent pre-handoff set. Percentages are relative to no VAD; 95\% intervals are [2465, 2842] ([78.45, 90.45]\%) for VAD and [277, 343] ([8.83, 10.91]\%) for PVAD.}
    \begin{tabular}{lccc}
        \toprule
        & No VAD & VAD & PVAD \\
        \midrule
        False interruptions
        & 3142 (100\%)
        & 2793 (88.9\%)
        & 310 (9.9\%) \\
        \bottomrule
    \end{tabular}
    \label{tab:full_duplex}
    \vspace{-0.7em}
\end{table}

Before transferring the integrated barge-in pipeline to the downstream product-validation team for broader deployment testing, we performed a final evaluation of the failure mode that motivated this work: false interruptions caused by residual assistant playback. Full-duplex voice assistants must reliably detect when a user interrupts while the assistant is speaking.
A deployable system must retain genuine target-user interruptions while rejecting target-absent false triggers. The LibriSpeech and real-user evaluations in Sections~\ref{sec:librispeech_evaluation} and~\ref{sec:real_user_evaluation} assess target-user retention through target-speaker discrimination, frame-level TPR, and target-event recall. This experiment provides complementary evidence by isolating target-absent false interruptions in the fully integrated pipeline. It is not intended to estimate end-to-end target-user recall.

\begin{figure}
    \centering
    \includegraphics[width=0.98\linewidth]{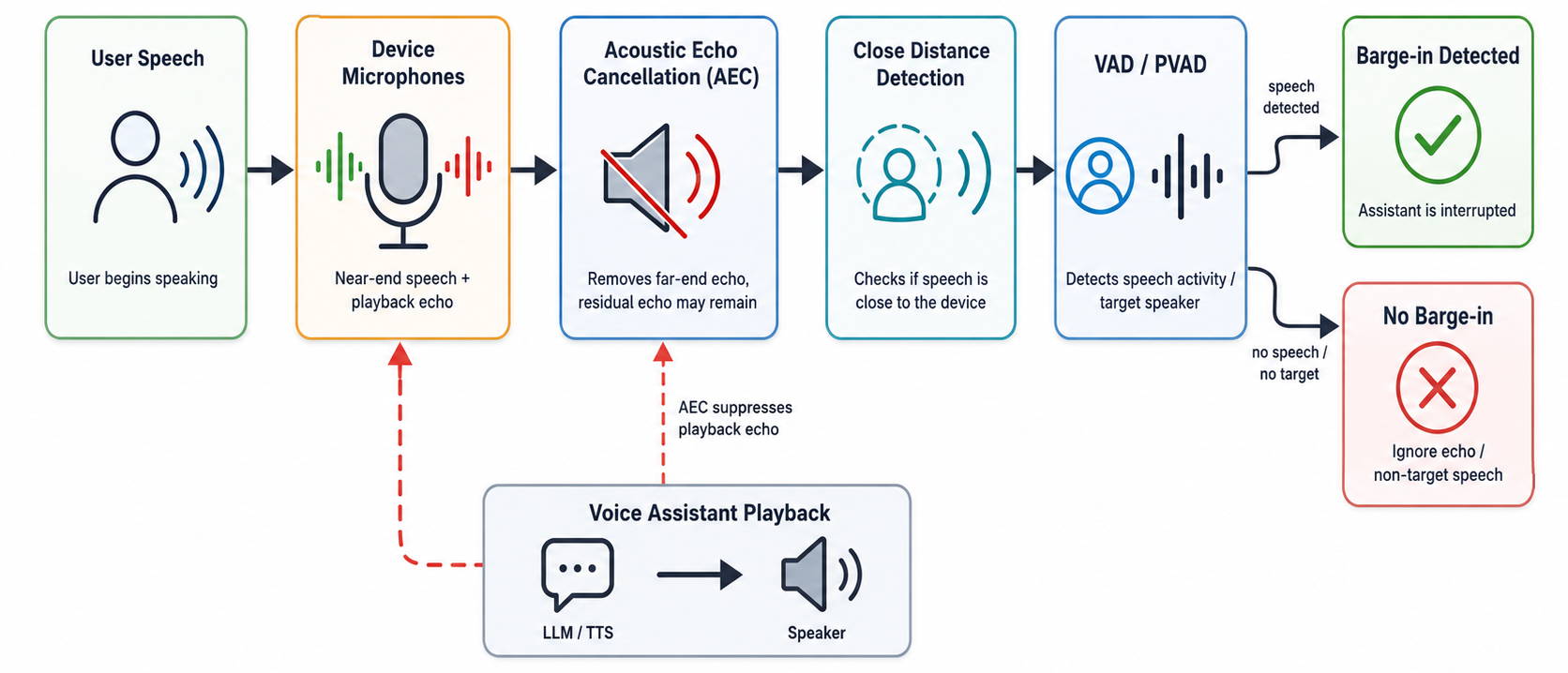}
    \caption{Barge-in detection pipeline: microphone input passes through AEC, close-distance detection, and VAD/PVAD to decide whether to interrupt the assistant. }
    \label{fig:full_duplex}
    \vspace{-1em}
\end{figure}

Our interruption detection pipeline, depicted in Figure~\ref{fig:full_duplex}, applies acoustic echo cancellation (AEC), followed by a close-distance detector that identifies speech activity when the device is within 15~cm of the user's mouth. A VAD then operates downstream. In practical mobile conditions, microphone occlusion, nonlinear echo residuals, and AEC failures cause the close-distance detector to pass residual assistant speech. A generic VAD cannot determine whether this speech belongs to the enrolled user, so we replace it with a PVAD while preserving the downstream interruption interface.

We construct a dedicated edge-case evaluation set from internal pre-release testing of the fully integrated pipeline, immediately before product handoff. The set contains segments that the upstream close-distance detector incorrectly classified as user interruptions.
It comprises 388 audio segments totaling more than 64 minutes. No target-user speech is present in these recordings; all detected speech originates from residual assistant playback. Hence, this dataset is intentionally restricted to target-absent examples, used only to measure false-interruption suppression.

Each segment is processed without VAD, with generic FSMN-VAD, and with FSMN-PVAD. Because the assistant voice differs from the enrolled user, every segment is a non-target trial. For PVAD evaluation, an enrolled profile from the speaker set in Section~\ref{sec:real_user_evaluation} is randomly assigned to each segment, providing coverage across enrollment identities.
Table~\ref{tab:full_duplex} reports the number and percentage of false interruption detections retained under each configuration.
The VAD provides only a modest reduction in false interruptions, while PVAD suppresses a great majority of false triggers.

\section{Conclusion}
\label{sec:concusion}

Voice-assistant pipelines typically rely on voice activity detection (VAD) as an initial gate. In full-duplex systems, a speaker-agnostic VAD confuses nearby speech or residual assistant playback with a user interruption event. Personalized VAD (PVAD) suppresses these errors by restricting activation to an enrolled user, but existing methods often require changes to the deployed VAD architecture. We presented HyWA, a weight-adaptation approach that converts a pretrained speaker-agnostic VAD into a personalized VAD without changing its inference pipeline or requiring per-user retraining. We demonstrated this approach using pretrained Alibaba's FSMN and NVIDIA's MarbleNet VAD backbones.

Our HyWA-based FSMN-PVAD significantly improved frame- and event-level performance relative to its matched speaker-agnostic VAD. These improvements also translated to an integrated full-duplex barge-in pipeline, where PVAD reduced the fraction of retained false interruptions from 88.9\% to 9.9\%. Therefore, architecture-preserving personalization substantially enhances the reliability of full-duplex voice interaction without modifying the inference pipeline.

\newpage 
\bibliography{mybib}

\clearpage
\appendix

\begin{figure*}[t]
    \centering
    \includegraphics[width=\textwidth]{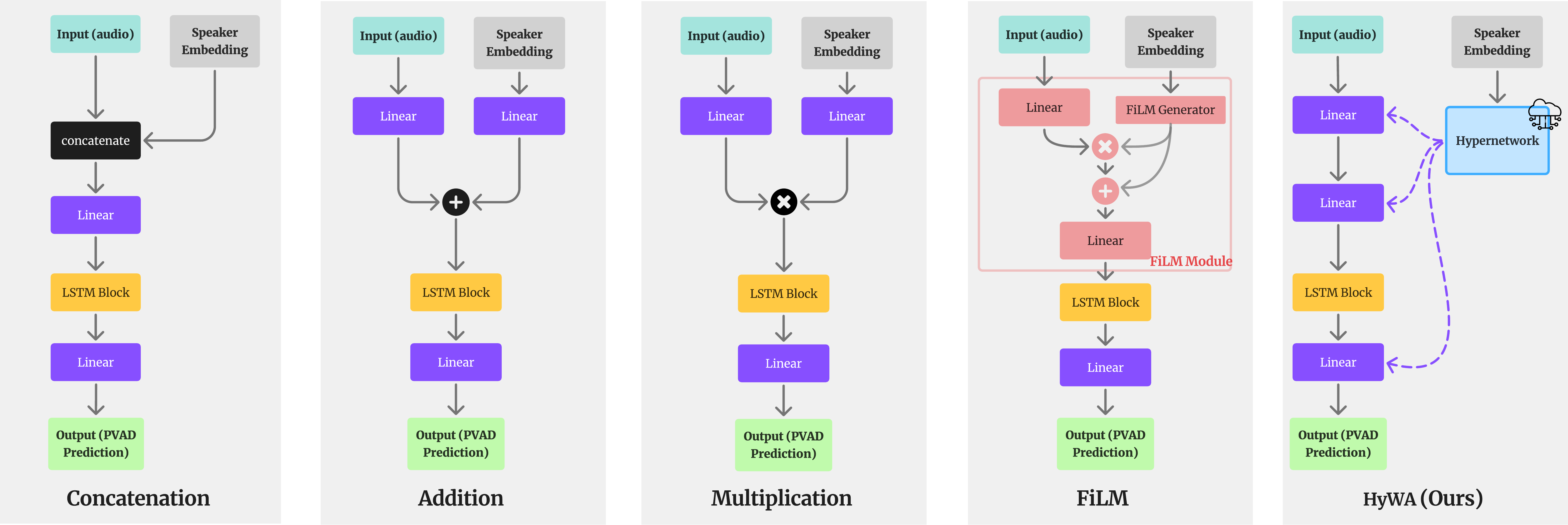}
    \caption{Speaker conditioning methods for PVAD. Each approach illustrates a distinct strategy for integrating speaker information with acoustic features to enable personalization.}
    \label{fig:speaker_cond}
\end{figure*}

\section{HyWA Proof of Concept}
\label{app:hywa_proof_of_concept}
This appendix reports the initial feasibility study of HyWA using a compact detector trained jointly with its hypernetwork. The binary Base reference in Table~\ref{tab:librispeech_results} uses the same compact backbone family with a binary target-presence output. Here, we use a ternary output to compare HyWA with alternative speaker-conditioning methods. The study examines whether speaker-conditioned weight generation can support personalized speech activity detection before applying the method to frozen pretrained VAD backbones. Because the compact system differs from the pretrained systems in model scale and task formulation, its results provide proof-of-concept evidence rather than a direct comparison with the pre-deployment models.

\subsection{Task Definition}
\label{app:proof_of_concept_task_def}
The proof-of-concept study uses the LibriSpeech mixture construction, speaker-disjoint partitions, noise augmentation, and room simulation described in Section~\ref{sec:librispeechdata}. The compact detector predicts three mutually exclusive frame-level classes: \{``non-speech~(ns)'', ``target-speaker speech~(tss)'', ``non-target-speaker speech~(ntss)''\}. This ternary label space is used consistently for HyWA and the alternative speaker-conditioning methods. In contrast, the binary Base reference and pretrained-backbone experiments combine ``ntss'' and ``ns'' into a single negative class. The ternary results are therefore evaluated under their own task definition.

\subsection{Model Architecture}

Our VAD model $\mathcal M$ is inspired by~\cite{SSL-pvad-baseline-journalVersion} and includes a 2-layer LSTM with 64 hidden units. We add a two-layer perceptron before the LSTM block and a single-layer perceptron after the LSTM block to boost personalization capacity, yielding an on-device VAD with $\approx 85$k parameters.
For speaker conditioning, following~\cite{SSL-pvad-baseline} we extract 256-dimensional speaker embeddings using the Resemblyzer VoiceEncoder~\cite{icassp2018_resemblyzerpaper}.
Each speaker embedding is computed from at least 3 seconds of enrollment audio in total, formed by concatenating one or more short utterances from that speaker.
The hypernetwork $\mathcal H$ takes this speaker embedding as input and outputs user-specific parameters $\Delta\mathbf w$ that personalize $\mathcal M$. We use a 4-layer perceptron with GeLU activations, normalization, and a skip connection at each layer, totaling $\approx 3.6$M parameters in the cloud.
We restrict personalization to linear layers to keep the design and analysis simple.

\subsection{Evaluation Metrics}

Following~\cite{personal_vad_1, SSL-pvad-baseline}, we evaluate the performance of the model by computing the average precision score (AP) for each class, with the mean average precision (mAP) serving as the primary evaluation metric. The mAP score is determined by taking the average of the AP scores across all classes.

\begin{table*}[!t]
\centering
\small
\setlength{\tabcolsep}{10pt}
\caption{Average precision (AP) scores for clean speech and speech in seen and unseen noise, averaged over all SNR levels. Values are percentages reported as mean (standard deviation) over five seeds. The best result in each condition and metric is shown in bold.}
\label{tab:base_case_results}
\begin{tabular}{cccccc}
\toprule
\multirow[c]{2}{*}{\shortstack{\\ \\ \textbf{Speaker} \textbf{Conditioning} \\ \textbf{Method}}} &
\multirow[c]{2}{*}{\shortstack{\\ \\ \\ \textbf{Scenario}}} &
\multicolumn{3}{c}{\textbf{AP (\%) $\uparrow$}} &
\multirow[c]{2}{*}{\shortstack{\\ \\ \\ \textbf{mAP (\%) $\uparrow$}}} \\
\cmidrule(lr){3-5}
& &
\shortstack{Non-Speech \\ \textbf{(ns)}} &
\shortstack{Target Speech \\ \textbf{(tss)}} &
\shortstack{Non-Target Speech \\ \textbf{(ntss)}} &
\\
\midrule

\multirow[c]{3}{*}{Concatenation}
  & Clean  & 93.7 (0.3) & 85.9 (1.0) & 89.5 (0.9) & 89.7 (0.7) \\
  & Seen Noise  & 83.5 (0.5) & 80.5 (0.7) & 85.2 (0.6) & 83.1 (0.6) \\
  & Unseen Noise & 83.9 (0.4) & 80.0 (1.1) & 84.4 (0.7) & 82.8 (0.7) \\
\midrule

\multirow[c]{3}{*}{Multiplication}
  & Clean  & 93.4 (0.2) & 85.9 (0.9) & 89.3 (0.5) & 89.6 (0.4) \\
  & Seen Noise  & 82.7 (0.6) & 81.2 (0.7) & 85.0 (0.5) & 83.0 (0.4) \\
  & Unseen Noise & 83.2 (0.5) & 80.1 (0.7) & 83.8 (0.6) & 82.3 (0.4) \\
\midrule

\multirow[c]{3}{*}{Addition}
  & Clean  & 93.8 (0.1) & 84.3 (1.4) & 88.5 (1.2) & 88.9 (0.9) \\
  & Seen Noise  & 83.8 (0.3) & 79.4 (1.0) & 84.6 (0.9) & 82.6 (0.7) \\
  & Unseen Noise & \textbf{84.0} (0.3) & 78.5 (1.6) & 83.6 (1.1) & 82.0 (0.9) \\
\midrule

\multirow[c]{3}{*}{FiLM}
  & Clean  & 93.9 (0.3) & 85.8 (0.5) & 89.3 (0.8) & 89.7 (0.4) \\
  & Seen Noise & 83.6 (0.6) & 81.8 (0.4) & 85.6 (0.6) & 83.7 (0.3) \\
  & Unseen Noise & 83.7 (0.6) & 80.7 (0.7) & 84.2 (0.8) & 82.9 (0.5) \\
\midrule

\multirow[c]{3}{*}{\makecell[c]{\textbf{HyWA}\\(Ours)}}
  & Clean & \textbf{94.1} (0.3) & \textbf{89.3} (0.6) & \textbf{91.3} (0.9) & \textbf{91.6} (0.4) \\
  & Seen Noise & \textbf{84.0} (0.8) & \textbf{85.6} (0.6) & \textbf{87.9} (0.5) & \textbf{85.9} (0.5) \\
  & Unseen Noise & \textbf{84.0} (0.8) & \textbf{85.4} (0.7) & \textbf{87.2} (0.5) & \textbf{85.5} (0.5) \\
\bottomrule
\end{tabular}
\end{table*}

\subsection{Results}

Table~\ref{tab:base_case_results} compares our proposed approach, HyWA, against four commonly used speaker-conditioning schemes for PVAD: (1) Concatenation, (2) Multiplication, (3) Addition, and (4) feature-wise linear modulation (FiLM); see Figure~\ref{fig:speaker_cond}.
For FiLM, note that we evaluated conditioning at multiple insertion sites, including pre- and post-LSTM block. We report the pre-LSTM FiLM configuration because it achieves 1.5\% higher mAP than post-LSTM in our setting, consistent with observations in~\cite{SSL-pvad-baseline-journalVersion}.

We consider three experimental scenarios. The first, ``Clean'', corresponds to a noise-free environment, where the dataset contains only clean speech signals. This setup allows us to examine the inherent capacity of each method without the confounding effect of noise.
The second, ``Seen Noise'', reflects a condition where the test set includes noise types observed during training. Finally, the ``Unseen Noise'' scenario evaluates generalization to novel acoustic conditions by incorporating noise types that were not observed during training.
For both noisy scenarios, we generate test utterances using the same SNR range and sampling procedure as in training, namely $-5$ to $20$ dB in $5$ dB increments. We report results averaged over SNR levels in Table~\ref{tab:base_case_results}, thereby providing a comprehensive measure of robustness.

Table~\ref{tab:base_case_results} indicates that HyWA improves AP and mAP over all four baselines across all scenarios.
In the ``Clean'' condition, our model attains the best AP and mAP, demonstrating strong target-speaker discrimination in ideal settings.
Under both ``Seen Noise'' and ``Unseen Noise'', our approach continues to outperform other competing methods, suggesting improved robustness and better target-speaker extraction.
Overall, these results indicate that HyWA is competitive with common speaker-conditioning alternatives and obtains the highest target-speech AP and mAP across the evaluated acoustic conditions. The deployment-oriented experiments separately evaluate its fixed-topology advantage on pretrained backbones.

\section{Pre-Deployment Model Configurations}
\label{app:predeployment_configs}

This appendix provides the architectural details of the hypernetworks that personalize the pretrained FSMN and MarbleNet VAD backbones. Both systems use a 256-dimensional, $\ell_2$-normalized enrollment embedding $\mathbf{e}_k$ produced by Resemblyzer VoiceEncoder~\cite{icassp2018_resemblyzerpaper}. The pretrained backbone parameters remain frozen during hypernetwork training.

For each adapted module $j$, an independent hypernetwork $\mathcal{H}_j$ maps the enrollment embedding to a full residual tensor:
\begin{equation}
    \Delta\mathbf{w}_{k,j}
    =
    \alpha\,\mathcal{H}_j(\mathbf{e}_k),
    \qquad
    \mathbf{w}_{k,j}
    =
    \mathbf{w}_j+\Delta\mathbf{w}_{k,j},
\end{equation}
where $\mathbf{w}_j$ denotes the corresponding frozen backbone parameters and $\alpha$ is a fixed, model-specific residual scale. When an adapted module contains a bias, the hypernetwork also generates a bias residual that is added to the pretrained bias. The generated tensors have the same shapes as their corresponding backbone parameters.

\subsection{FSMN-PVAD Configuration}
\label{app:fsmn_pvad_config}

FSMN-PVAD uses six independent hypernetwork blocks to adapt six parameter groups: the four $250\times128$ affine layers in the FSMN blocks, the $140\times250$ output affine layer, and the $2\times140$ binary classification head. Each block contains an input LayerNorm operation, a $256$-to-$512$ linear projection, a GELU activation, dropout with probability $0.1$, and a layer-specific linear emitter, i.e. weight generator. The emitted values are scaled by $\alpha=0.02$ and reshaped to match the corresponding weight and bias tensors.

Collectively, the six hypernetwork blocks contain 85.14 million trainable parameters and generate 164,422 user-specific residual parameters for each enrollment embedding.

\subsection{MarbleNet-PVAD Configuration}
\label{app:marblenet_pvad_config}

MarbleNet-PVAD uses thirteen independent hypernetwork blocks to adapt twelve point-wise convolutional layers and the $128$-to-$2$ linear decoder. Each hypernetwork contains two hidden linear layers with dimensions $256$-to-$512$ and $512$-to-$512$, together with GELU activations, hidden-layer normalization, and dropout with probability $0.2$. A layer-specific linear emitter generates the residual weights, i.e. parameters, for its corresponding MarbleNet parameter group. The emitted values are scaled by $\alpha=0.1$ before being reshaped and added to the pretrained parameters.

Collectively, the thirteen hypernetwork blocks contain 46.26 million trainable parameters and generate 80,130 user-specific residual parameters for each enrollment embedding.

\begin{table}[t]
    \centering
    \caption{Hypernetwork configurations for the pretrained VAD backbones.
    Generated parameters are the user-specific residuals loaded into each
    backbone, and trainable parameters denote the complete set of hypernetwork parameters.}
    \label{tab:predeployment_hypernetwork_configs}

    \begingroup
    \footnotesize
    \setlength{\tabcolsep}{3pt}
    \renewcommand{\arraystretch}{1.08}

    \begin{tabularx}{\columnwidth}{
        @{}
        >{\raggedright\arraybackslash}X
        >{\centering\arraybackslash}p{0.22\columnwidth}
        >{\centering\arraybackslash}p{0.22\columnwidth}
        @{}
    }
        \toprule
        & \multicolumn{2}{c}{Backbone} \\
        \cmidrule(lr){2-3}
        Configuration
            & MarbleNet
            & FSMN \\
        \midrule
        Enrollment dimension
            & 256
            & 256 \\
        Independent hypernetwork blocks
            & 13
            & 6 \\
        Hidden dimensions
            & 512, 512
            & 512 \\
        Dropout probability
            & 0.2
            & 0.1 \\
        Residual scale $\alpha$
            & 0.1
            & 0.02 \\
        Generated parameters
            & 80,130
            & 164,422 \\
        Trainable parameters
            & 46.26M
            & 85.14M \\
        \bottomrule
    \end{tabularx}
    \endgroup
\end{table}

\subsection{Personalization}
\label{app:parameter_application}

Both configurations generate full residual tensors rather than low-rank or factorized updates. The scaled residuals are added directly to the corresponding pretrained weights and biases; no gating, clipping, output normalization, or additional nonlinear transformation is applied. Once the personalized parameters have been generated and loaded, speech-time inference uses the existing backbone topology and audio interface. Enrolling or switching users therefore requires generating and loading a different residual configuration, without gradient-based optimization of the VAD for the individual user.

\section{Operating-Point Sensitivity}
\label{app:fpr_sensitivity}

We repeat the real-user operating-point analysis at target development-set frame FPRs of 0.05, 0.10, and 0.20 without retraining or rerunning model inference. Each model and PVAD seed is calibrated independently on the same eight development groups. Its selected threshold is then frozen before evaluation on the 18 held-out test groups. The target development FPR is therefore a calibration objective, not the achieved test FPR reported in Table~\ref{tab:fsmn_fpr_sensitivity}. Frame metrics retain the common 20-ms grid, whereas activation regions and target events retain each model's native valid timeline, the strictly-less-than-100-ms merging rule, and the 100-ms minimum-overlap rule.

\begin{table*}[t]
\centering
\small
\setlength{\tabcolsep}{4pt}
\caption{Operating-point sensitivity for FSMN-VAD and FSMN-PVAD. Thresholds target development-set frame FPR and are frozen for held-out test evaluation. Values are pooled estimates (10,000-replicate cluster-bootstrap SE); PVAD estimates average five fixed training seeds.}
\label{tab:fsmn_fpr_sensitivity}
\begin{tabular}{c l c c c c c c}
\toprule
Dev FPR & System & Test FPR (\%) $\downarrow$ & FP s/h $\downarrow$ & TPR (\%) $\uparrow$ & BA (\%) $\uparrow$ & FA/h $\downarrow$ & Event recall (\%) $\uparrow$ \\
\midrule
0.05 & VAD & 5.7 (0.2) & 203.5 (7.3) & 15.8 (0.6) & 55.1 (0.3) & 1357.3 (51.7) & 63.2 (1.7) \\
0.05 & PVAD & 5.0 (0.2) & 180.2 (8.2) & 33.3 (1.0) & 64.1 (0.5) & 479.3 (21.4) & 73.4 (1.4) \\
\addlinespace
0.10 & VAD & 11.5 (0.4) & 412.8 (13.3) & 27.7 (0.9) & 58.1 (0.4) & 1714.7 (51.5) & 75.8 (1.4) \\
0.10 & PVAD & 9.7 (0.5) & 350.5 (18.1) & 48.6 (1.4) & 69.4 (0.5) & 778.4 (34.2) & 86.4 (1.4) \\
\addlinespace
0.20 & VAD & 22.5 (0.6) & 808.6 (23.3) & 46.6 (1.1) & 62.1 (0.4) & 1759.0 (57.0) & 89.7 (0.9) \\
0.20 & PVAD & 19.3 (1.1) & 696.6 (39.1) & 67.8 (1.8) & 74.2 (0.5) & 1170.4 (25.8) & 94.8 (1.0) \\
\bottomrule
\end{tabular}
\end{table*}

\begin{table*}[t]
\centering
\small
\setlength{\tabcolsep}{4pt}
\caption{Operating-point sensitivity for MarbleNet-VAD and MarbleNet-PVAD. Thresholds target development-set frame FPR and are frozen for held-out test evaluation. Values are pooled estimates (10,000-replicate cluster-bootstrap SE); PVAD estimates average five fixed training seeds.}
\label{tab:marblenet_fpr_sensitivity}
\begin{tabular}{c l c c c c c c}
\toprule
Dev FPR & System & Test FPR (\%) $\downarrow$ & FP s/h $\downarrow$ & TPR (\%) $\uparrow$ & BA (\%) $\uparrow$ & FA/h $\downarrow$ & Event recall (\%) $\uparrow$ \\
\midrule
0.05 & VAD & 5.7 (0.3) & 205.4 (9.0) & 16.9 (0.9) & 55.6 (0.4) & 247.0 (12.4) & 37.4 (1.7) \\
0.05 & PVAD & 4.7 (0.2) & 167.9 (7.1) & 22.5 (1.0) & 58.9 (0.5) & 165.3 (6.1) & 40.9 (1.5) \\
\addlinespace
0.10 & VAD & 11.0 (0.4) & 396.5 (15.5) & 28.9 (1.3) & 58.9 (0.5) & 327.0 (13.6) & 52.8 (2.0) \\
0.10 & PVAD & 9.2 (0.3) & 331.1 (12.2) & 35.0 (1.2) & 62.9 (0.6) & 274.7 (9.1) & 55.8 (1.3) \\
\addlinespace
0.20 & VAD & 21.4 (0.8) & 771.5 (28.3) & 49.1 (1.6) & 63.8 (0.5) & 382.0 (14.4) & 73.3 (1.9) \\
0.20 & PVAD & 18.3 (0.6) & 657.0 (20.8) & 53.4 (1.2) & 67.6 (0.6) & 418.3 (12.7) & 74.1 (0.8) \\
\bottomrule
\end{tabular}
\end{table*}

For the deployment-candidate FSMN pair, the qualitative trade-off is consistent across these three analysis points: FSMN-PVAD has lower test FPR, FP seconds per target-absent hour, and false activations per recording hour, while retaining higher target-speech TPR, balanced accuracy, and target-event recall than FSMN-VAD. Paired 95\% cluster-bootstrap intervals exclude zero for all six PVAD-minus-VAD effects at every evaluated point. This sensitivity analysis does not designate a product operating point; that choice requires joint validation with the complete application pipeline.

\section{Training and Reproducibility Details}
\label{app:training_details}

Table~\ref{tab:training_details} records the optimization settings from the final configurations and trainer implementations. A batch denotes mixture utterances: FSMN accumulates four two-utterance minibatches per update, whereas the compact and MarbleNet systems update after 64 utterances. The compact proof-of-concept jointly optimizes its from-scratch backbone and hypernetwork. For FSMN-PVAD and MarbleNet-PVAD, the pretrained backbone remains frozen.

\begin{table*}[t]
  \centering
  \small
  \setlength{\tabcolsep}{4pt}
  \caption{Training settings.}
  \label{tab:training_details}
  \begin{tabular}{c c c c c}
    \toprule
    System & Optimizer & LR / schedule & Batch & Maximum / stopping \\
    \midrule
    Compact proof of concept
      & AdamW
      & $10^{-3}$ / warm-restart cosine
      & 64
      & 100 epochs / patience 5 \\
    FSMN-PVAD
      & AdamW, wd $10^{-4}$
      & $10^{-3}$ / constant
      & $2\times4=8$
      & 50 epochs / patience 10 \\
    MarbleNet-PVAD
      & AdamW, wd $10^{-4}$
      & $10^{-3}$ / constant
      & 64
      & 100 epochs / patience 10 \\
    \bottomrule
  \end{tabular}
\end{table*}

Ternary version of compact system uses unweighted three-class cross-entropy and clips hypernetwork gradients to norm 1.0. FSMN-PVAD and MarbleNet-PVAD use weighted binary cross-entropy with target-class weight 2.0. Both clip trainable gradients to norm 5.0. The compact frontend uses 16-kHz audio, 40 mel bands, a 25-ms window, and a 10-ms hop. The pretrained systems consume 16-kHz waveforms through their restored FSMN/MarbleNet frontends. Training augmentation adds free-sound noise from $-5$ to $20$ dB and room impulse responses, each with probability 0.5. The FSMN VAD is derived from FunASR 1.3.14.

\section{Bootstrap Uncertainty Estimation}
\label{app:event_bootstrap}

For comparisons of the VAD and PVAD results reported in Tables~\ref{tab:application_fp_reduction} and~\ref{tab:application_event_results}, we define the paired difference as
$\Delta=\mathrm{PVAD}-\mathrm{VAD}$.
We estimate uncertainty using 10,000 paired cluster-bootstrap replicates over the 18 held-out evaluation groups. In each replicate, groups are sampled with replacement, and the resulting group multiplicities are applied identically to all systems being compared. Each pooled metric is recomputed from its resampled aggregate numerator and denominator rather than by averaging group-level metric values.
For PVAD, the pooled metric is recomputed separately for each of the five fixed training seeds and then averaged across seeds within each replicate. The training seeds are not resampled. The resulting uncertainty estimates therefore characterize variation across evaluation groups conditional on the five evaluated PVAD models; they do not capture variability attributable to retraining with different random seeds.
Bootstrap standard errors reported in the tables are obtained from these replicates. For paired PVAD--VAD comparisons, we additionally report point-wise 95\% percentile-bootstrap intervals for $\Delta$.

\end{document}